# High Capacitive Energy Density in layered 2D Nanomaterial based Polymer Dielectric Films

Maninderjeet Singh[1,†], Priyanka Das[2,†], Pabitra Narayan Samanta[2], Sumit Bera[2], Ruskshan Thantirige[2], Brian Shook[2], Roshanak Nejat[3], Banarji Behera[4], Qiqi Zhang[2], Qilin Dai[2], Avijit Pramanik[2], Paresh Ray[2], Dharmaraj Raghavan[5], Jerzy Leszczysnki[2], Alamgir Karim[1,*], and Nihar R. Pradhan[2,*]

[1]*Department of Chemical & Biomolecular Engineering, University of Houston, Houston, TX 77204, USA,*

[2]*Department of Chemistry, Physics & Atmospheric Sciences, Jackson State University, Jackson, MS 39217, USA,*

[3]*Materials Engineering Program, University of Houston, Houston, TX 77204, USA,*

[4]*School of Physics, Sambalpur University, Jyoti Vihar, Burla, Sambalpur, Odisha 768019, India,*

[5]*Department of Chemistry, Howard University, Washington DC 20059, USA*

*E-mail: nihar.r.pradhan@jsums.edu, akarim3@central.uh.edu

† Authors contributed equally.

**Abstract**

Dielectric capacitors are critical components in electronics and energy storage devices. The polymer-based dielectric capacitors have the advantages of device flexibility, fast charge-discharge rates, low loss, and graceful failure. Elevating the use of polymeric dielectric capacitors for advanced energy applications such as electric vehicles (EVs) however, requires significant enhancement of their energy densities. Here, we report a polymer thin film heterostructure-based capacitor of poly(vinylidene fluoride)/poly(methyl methacrylate) with stratified 2D nanofillers (Mica or *h*-BN nanosheets) (PVDF/PMMA-2D fillers/PVDF), that shows enhanced permittivity, high dielectric strength and an ultra-high energy density of ≈ 75 J/cm$^3$ with efficiency over 79%. Density functional theory calculations verify the observed permittivity enhancement. This approach of using oriented 2D nanofillers-based polymer heterostructure composites is expected to be versatile for designing high energy density thin film polymeric dielectric capacitors for myriads of applications.

**One-Sentence Summary:** A layered thin film dielectric capacitor based on PVDF, PMMA, and exfoliated 2D nanofillers (Mica or *h*-BN) exhibits an ultra-high energy density of ≈ 75 J/cm$^3$.

## Table of contents

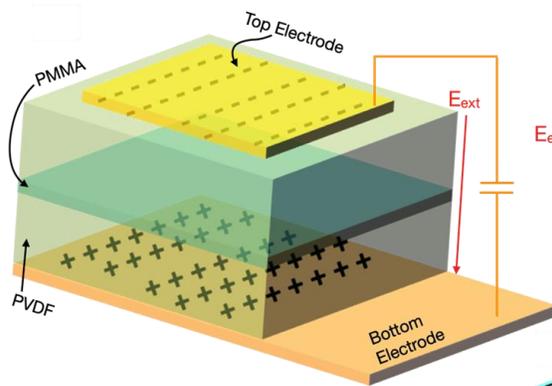
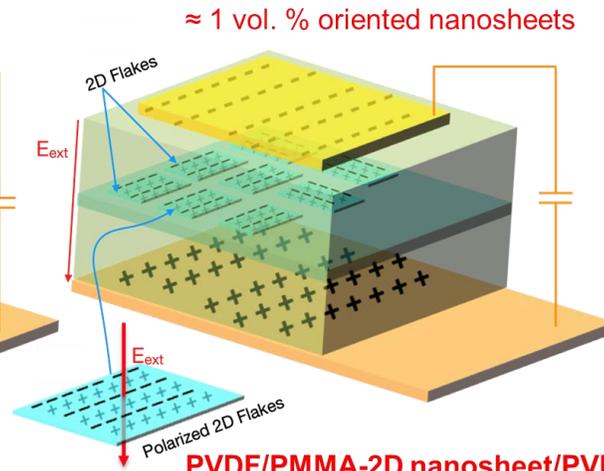

**PVDF/PMMA/PVDF**
**Dielectric Strength ($E_{BD}$) – 535 MV/m**
**Max. Energy Density ($U_D$) – 4.5 J/cm³**

**PVDF/PMMA-2D nanosheet/PVDF**
**Dielectric Strength ($E_{BD}$) – 1296 MV/m**
**Max. Energy Density ($U_D$) – 75 J/cm³**

**Introduction**

High energy density and high-power density energy storage dielectric capacitors will play an important role in enabling reliable power supply as decarbonization drives the world to shift towards intermittent renewable energy sources [1]. However, the energy density of existing dielectric capacitors remains limited as compared to their electrochemical counterparts. The energy stored per unit volume in a dielectric material can be expressed in terms of electric displacement ($D$), electric field ($E$), residual electric displacement ($D_r$), and maximum displacement ($D_{max}$) [2] as

$$U = \int_{D_r}^{D_{max}} E dD$$

The electric displacement is $D \propto \varepsilon$, where $\varepsilon$ is the permittivity. The maximum energy stored in the dielectric material is dictated by the maximum electric field that the material can withstand and is referred to as the dielectric strength ($E_{BD}$) of the material. Thus, to increase the maximum energy density of the materials, both $\varepsilon$ and $E_{BD}$ need to be increased [3,4]. Polymer-based dielectric capacitors have a lower loss, tunable dielectric strength with nanofillers, flexibility, and easy fabricability for free-standing capacitors as compared to ceramic dielectrics [5,6]. However, due to the low permittivity and $E_{BD}$ of the polymeric dielectrics, their energy densities are typically low, e.g. Biaxially Oriented Polypropylene (BOPP) polymer has $U \approx$ 2-3 J/cm$^3$, but still has a wide range of applications. Embedding nanofillers in a polymer matrix can enhance their permittivity but the dielectric strength is limited due to a random distribution of the fillers, resulting in low to moderate energy densities. Furthermore, the nanoparticle aggregation and air voids in these nanocomposites decrease dielectric strength, thereby negatively impacting the energy stored in the dielectrics[7–9]. Recently, 2D nanofiller-based polymer nanocomposites[5,10] have shown potential for use as relatively higher energy density energy storage devices such as Ca$_2$Nb$_3$O$_{10}$ based nanosheets in polyvinylidene fluoride (PVDF) matrix have shown energy densities of 36 J/cm$^3$ with efficiencies of 60% [11]. However, the energy densities and efficiencies of these nanocomposites remain low for practical application [12]. Theoretical and simulated studies have predicted that the alignment control of 2D nanofillers can enhance the energy density of polymeric capacitors when nanosheets are perpendicularly oriented to the applied electric field, can delay the dielectric breakdown by reducing the electrical treeing effect and enhance the polarization field by adding interfacial dipoles [13], however, it is difficult to fully orient these 2D nanofillers in polymer matrices in a particular direction using dispersion methods.

In this report, we designed layered polymer nanocomposites based on oriented 2D nanofillers in PVDF as the host polymer and mechanically exfoliated flakes (Fig.1)) of layered 2D materials (2D-Mica or hexagonal Boron Nitride (*h*-BN) crystals) as the nanofillers in a heterostructure geometry (Fig. 1A). The heterostructure referred to here is poly(vinylidene fluoride)/poly(methyl methacrylate)-oriented 2D Mica (or *h*-BN) nanosheets/poly(vinylidene fluoride) (PVDF/PMMA-2D Mica or *h*-BN/PVDF) stacking layers, wherein layered bulk crystals are mechanically exfoliated to thinner layers on to SiO$_2$/Si substrate (Fig 1A) and transferred onto the thin layer of a PVDF film (bottom PVDF) using PMMA and subsequently sandwiched with another thin layer of top PVDF film (Fig 1B).

In this heterostructure capacitor design, the 2D surfaces of all the inorganic nanofillers are parallel to the capacitor electrodes, so that the externally applied electric field is perpendicular to the plane of 2D nanofiller sheets. We observe a significantly high enhancement of dielectric constant ($\varepsilon$) ($\Delta\varepsilon$ ≈ 60-100 %, depending on filler fraction (Fig. S1 in SI)) as well as breakdown strength ($\Delta E_{BD}$ ≈ 142 %) and an ultra-high capacitive energy density of ≈ 75 J/cm$^3$ with an efficiency over 79 %, using ≈1 vol % 2D Mica as fillers, which is highest among any reported polymer or polymer nanocomposite dielectrics [6,14–19] (higher by a factor of 2) for ≈ 0.6 µm thick polymer dielectric film (SI Fig. S3). As shown in simulations, this is due to the highly parallel nature of Mica nanosheets with high surface coverage (≈ 70%). Notably, the volume fraction of the nanofillers in the film is very low compared to blend nanocomposites, as the nanofillers are only present in a single layer between two layers of PVDF. Furthermore, we demonstrate that the effect of nanofiller orientation control to develop ultra-high energy density polymeric dielectrics is versatile by showing that $h$-BN based PVDF/PMMA-$h$-BN/PVDF heterostructures also show a high energy density of ≈ 50 J/cm$^3$ with an efficiency of ≈ 52%, which is higher than other reported dielectric capacitors expect the Mica based heterostructure capacitors studied here.

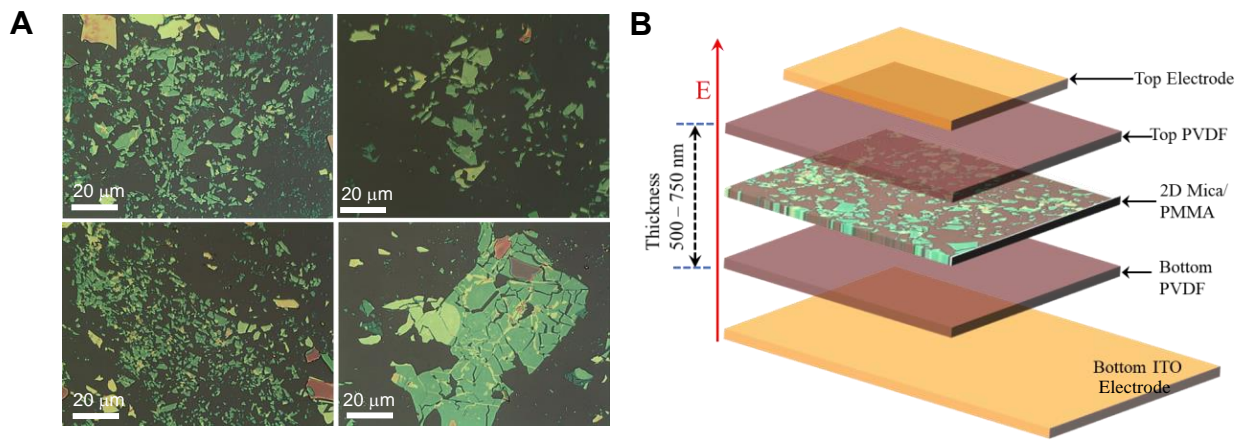

**Figure 1. Schematic of geometrical stacking of oriented 2D nanomaterial-based polymer heterostructure:** (A) Exfoliated mica flakes on the Si/SiO$_2$ substrate from a bulk crystal using mechanical exfoliation technique using scotch tape. Exfoliated flakes show large nonuniformity across the substrate. High-density flakes can be seen in a small area of the substrate. Four images taken from different sections of the wafer show the varying density of the flakes. (B) Schematic representation of PVDF/2D Mica-PMMA/PVDF heterostructure. The thickness of the heterostructure from the bottom to the top PVDF layer ranges from 500 nm to 750 nm. The middle 2D Mica layer is supported by a thin PMMA layer of thickness ≈ 150 nm.

## Results and Discussion

**Fabrication of Heterostructure Capacitor**

The schematic design of the heterostructure device consists of layers of PVDF and Mica (supported by the PMMA layer) as shown in **Fig. 2**(A-H). The Mica nanosheets were exfoliated using scotch tape through mechanical exfoliation technique and then transferred onto 285 nm $SiO_2$ deposited on Si substrate in such a way that half of the $SiO_2$ wafer remained empty, and half of the wafer was covered with transferred Mica flakes (Schematic of **Fig. 2A**). The empty half of the substrate was used as a reference capacitor and Mica covered half area used as the 2D filler interfaced capacitor. A Raman spectroscopy study was used on a thick layer of Mica crystal to verify the Mica crystal (**Fig. S2**). After the transfer process, we visually looked at the 2D Mica flakes under the optical microscope to verify the quality of distribution of the exfoliated crystals on the $Si/SiO_2$ substrate. The thickness of the 2D Mica flakes varies from a single layer to several atomic layers but the average thickness of most of the flakes obtained from AFM measurements shows 20 nm thickness (See **Fig. S3** (a-g)). The optical image shown in **Fig. 1** indicates the size and distribution of the flakes are random and nonuniform and the density of flakes varies across the exfoliated area of the substrate. A thin layer of PMMA was coated on the top of the exfoliated Mica sheets by spin coating a PMMA solution at 1600 rpm followed by annealing at 110 $^o$C for 2 minutes, providing good adhesion of the flakes to PMMA.

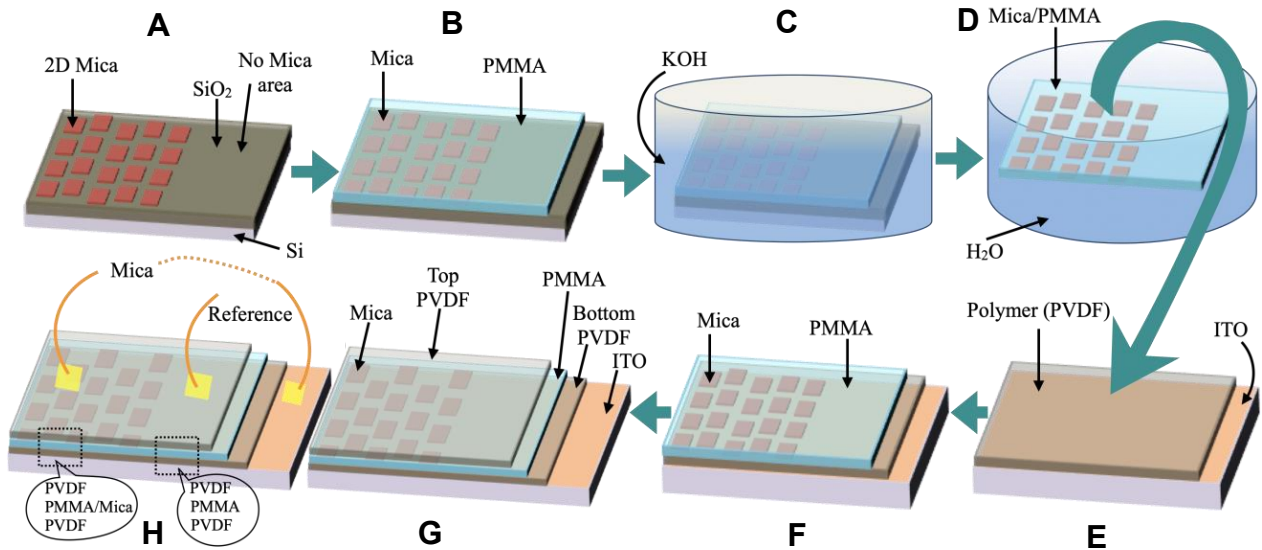

**Figure 2. Fabrication of oriented 2D nanomaterial-based polymer heterostructures:** (A) Mechanically exfoliated 2D Mica flakes (red squares) are transferred onto a clean SiO$_2$/Si substrate. On purpose, a section of the SiO$_2$/Si wafer was kept clean without transferring any 2D flakes. (B) The whole wafer was covered with PMMA polymer using the spin coating technique and baked at 100 $^o$C for 2 minutes. (C) The whole wafer (PMMA/2D-Mica/SiO$_2$/Si) was dipped inside the KOH solution. This etches the SiO$_2$ layer and releases the PMMA holding the 2D Mica layers to the solvent. (D) PMMA holding 2d-layered Mica crystals was later fished using a clean glass slide and transferred into the H$_2$O to remove any excess KOH solvent from the PMMA/2D-Mica layers. (E) The 2D-Mica/PMMA layer was then transferred onto the spin-coated PVDF film (bottom PVDF) on the ITO glass slides and the resulting device contained layers of ITO/PVDF/Mica-PMMA as shown in the schematic (F). (G) Finally, the top polymer layer (PVDF) was coated to complete the PVDF/2D Mica-PMMA/PVDF heterostructure device. (H) Shows the schematic of the final device used for dielectric measurements. The reference as well as mica-interfaced areas of the sample is labeled on the edge of the schematic.

Finally, this PMMA layer with 2D-Mica nanosheets was etched from the Si/SiO$_2$ substrate using SiO$_2$ etchant, and KOH solvent and transferred onto the PVDF coated layer on ITO substrate to obtain an ITO/PVDF/2D Mica-PMMA heterostructure [**Fig 2 (E-G)**]. After subsequent drying in vacuum, a top thin layer of PVDF was coated on to obtain the final PVDF/2D Mica-PMMA /PVDF heterostructure on the ITO as shown in **Fig. 2 (G)**. Half of the substrate yields a reference capacitor, PVDF/PMMA /PVDF, and the other half yields a PVDF/2D Mica-PMMA /PVDF heterostructure capacitor shown in **Fig. 2(H)**. In our fabricated capacitor devices, we measured the thickness of the PVDF/PMMA-2D Mica/PVDF heterostructure and found that the average thickness is ~500 nm ± 50nm (for the S6 sample in **Fig S3**). The non-uniformity arises from the manual transfer process of 2D Mica/PMMA layers. We kept the thickness of the polymer layers as low as possible to observe the maximum effect of 2D fillers on dielectric properties from the heterostructure assembly of the device as well as to limit the leakage or dielectric loss. In this vertically stacked geometry, the 2D surfaces of Mica flakes are parallel to the electrodes or perpendicular to the externally applied electric field $\vec{E}$ as indicated in **Fig. 1 (B)**. We observe that the 2D Mica nanofillers are intact in the final heterostructures using scanning electron microscopy (SEM). The SEM images on final heterostructures are shown in **Figure S5 (k)** in the SI.

**Permittivity Enhancement in Polymer-2D Layered Mica Heterostructures**

Figure 3 (A) and (B) display the dielectric constant of reference (PVDF/PMMA/PVDF) and stratified Mica interfaced capacitors (PVDF/PMMA-2D Mica/ PVDF) respectively as a function of temperature and frequency (SI Fig. S4). The reference capacitor shows a dielectric constant ($\varepsilon$) = 11-13.5 at room temperature, similar as compared to pristine PVDF. The 2D mica nanofiller-based heterostructures capacitor shows $\varepsilon$ = 19-22 at room temperature as a function of frequency. The permittivity of 2D Mica-based heterostructures is 60-75% higher than the reference heterostructure samples as shown in Figure 3(C). The enhancement, $\Delta\varepsilon$ of oriented 2D Mica-based heterostructures is much higher than the expected volumetric contribution of permittivity by the Mica nanofillers given that the filler volume fraction is very small ($\approx$ 1 vol%, see SI) and Mica

permittivity ranges from 6-9 [20]. We believe that the highly enhanced permittivity of 2D Mica-based samples stems from the additional interfacial diploes along the 2D plane due to the perpendicular (to the electric field) orientation of nanofillers. The large surface area of the 2D fillers provides ultra-large interfacial area per unit volume with the polymers where each small 2D flake acts as a micro-capacitor as shown in Fig. 3(F). The enhancement values of permittivity, $\Delta\varepsilon$ varies (SI Fig S7 and S10) with the density of the filler content and we also observed low enhancement in some area of the capacitor presumably from low 2D filler contain capacitor (SI Fig. S7 and S10). Interestingly, the loss tangent of the 2D nanofiller-based capacitor is comparable to the control reference capacitor at low frequency as shown in Figures 1D and 1E, which demonstrates the capability of the heterostructured architectures for developing high permittivity low loss polymeric dielectrics, especially since PVDF has relatively higher loss tangent and fillers tend to amplify this effect further. As such, this strategy would be useful for other low-loss dielectrics including commercially available BOPP. Typically, such enhancements in permittivity of polymers are obtained by high loadings of randomly dispersed inorganic nanofillers ($\geq$ 10 vol%), which results in significantly high losses as well [6]. A histogram plot of dielectric from 5 capacitors is presented in SI Fig. S1 and S7 shows as high as 150% enhancement (at higher filler fraction) from the heterostructure of PVDF/PMMA-2D/PVDF capacitors. To ensure that the permittivity calculations from asymmetric capacitors are accurate, we tested the control PVDF samples with symmetric and asymmetric top and bottom electrodes and found that the permittivities from asymmetric capacitors (by using smaller electrode area as capacitor area) and symmetric capacitors are the same. The detailed calculations from control symmetric and asymmetric capacitors along with the capacitor images are shown in SI Fig. S15.

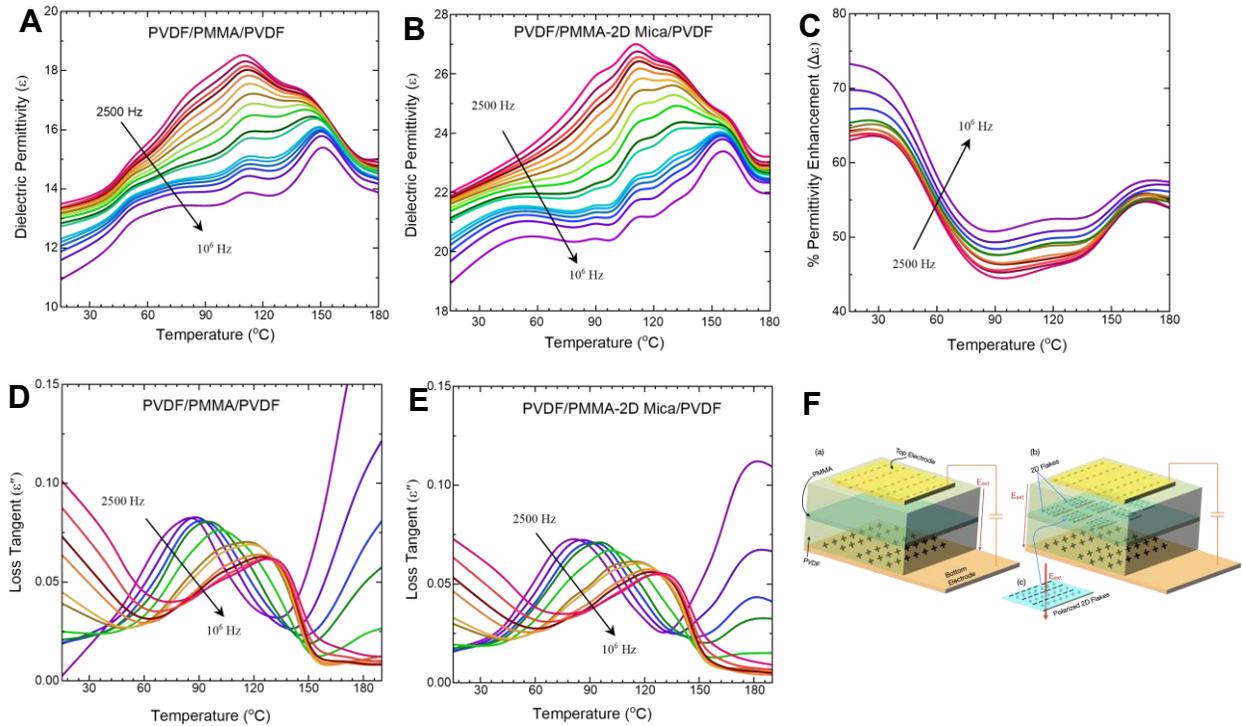

**Fig. 3. Enhanced permittivity and low loss in oriented 2D nanomaterial-based polymer heterostructures:** (A) and (B) Temperature ($T$) and Frequency (in Hz) dependent dielectric constant ($\varepsilon$) of PVDF/PMMA/PVDF and PVDF/2D Mica-PMMA/PVDF (Mica interfaced) heterostructure capacitors respectively. The thickness of the capacitor was measured as 500 nm (SI Fig S3). (C) Enhancement in $\varepsilon$ ($\Delta\varepsilon$) of Mica interfaced capacitor compared to the reference as a function of frequency. (D) and (E) show the dielectric loss for reference and Mica interfaced capacitors respectively. (F) Schematic showing origins of higher dielectric constant due to higher nanofiller-polymer interfacial polarization.

**Verification of Permittivity Enhancement using Density Functional Theory**

Given that we observe non-intuitive permittivity enhancement in oriented 2D Mica nanofiller samples at very low fractions, we perform density functional theory (DFT) calculations to unravel the physics of this phenomenon. The frequency-dependent complex dielectric function within the framework of Kubo-Greenwood formalism, in conjunction with the Kohn-Sham DFT and the Kramers-Krönig transformation was analyzed to understand the interface effect [21,22]. The dielectric properties of the isolated nanomaterials and their interfaces with the polymer matrix are estimated in response to the alteration of electric field (E) polarization. The predicted real parts of the dielectric tensor components as a function of the incident charge carrier energy for the Mica, PVDF, PMMA, and PVDF-Mica are portrayed in Figures 4 (A-D). The molecules of the polymers are favorably aligned parallel to the 2D plane at the interface, which increases the electrical polarization along the plane due to polarization along the polymer chain. A closer inspection of the dielectric tensor components of the pristine Mica and Mica-PVDF interface reveals that the limiting value of the $\varepsilon_{xx}$ enhances from 2.33 to 5.45, i.e., the dielectric constant of the mica is augmented by > 100% due to the attachment of the PVDF matrix. Furthermore, as evident from the threshold values of $\varepsilon_{xx}$ components of pristine PVDF and mica-PVDF composite, the dielectric constant of PVDF also rises from 2.35 to 5.45 in the mica-PVDF. Alternatively, the obtained results certainly manifest the essence of the in-plane orientation of the PVDF chain axis to maximize the dielectric response of the heterostructure. The maximum static dielectric constant of the Mica-PVDF is 75% higher than the predicted limiting value of the dielectric constant of pristine muscovite mica crystal and 130% higher than the pristine PVDF polymer, which is consistent with the experimental net dielectric measurements in heterostructure capacitor.

To explore the subtle interplay of Mica-Polymer interactions toward the improvement of dielectric response, we subsequently analyzed the electrostatic difference potential (EDP, $\Delta V_E$) across the interface which is characterized by the divergence between the electrostatic potentials derived from the self-consistent valence electron density and the superposition of atomic valence electronic densities. The plane-averaged EDP ($\Delta V_E$) along the z-direction (normal to the heterostructure) as obtained by solving Poisson's equation for electrostatics is depicted in Figure 4E. The calculated $\Delta V_E$ values manifest that some electrons transfer from the mica surface to the PVDF matrix at the interface which gives rise to the interfacial dipole, contributing to an improved dielectric response in the mica-PVDF heterostructure. The robust intermolecular interaction between the Mica layer and the PVDF matrix is evident by the charge density difference contour plot as shown in Fig. 4F. Figure 4G describes the 2D electron localization function (ELF) profile map across the interaction region of the Mica-PVDF composite. The formation of chemical bonding around the interaction sites is substantiated by the higher values of the ELF index ($> 0.7$), and consequently, the optical functions of the Mica-PVDF heterostructure including dielectric constant, refractive index (SI section Fig. S12-S13), etc. are significantly modulated by the high-polarity interface.

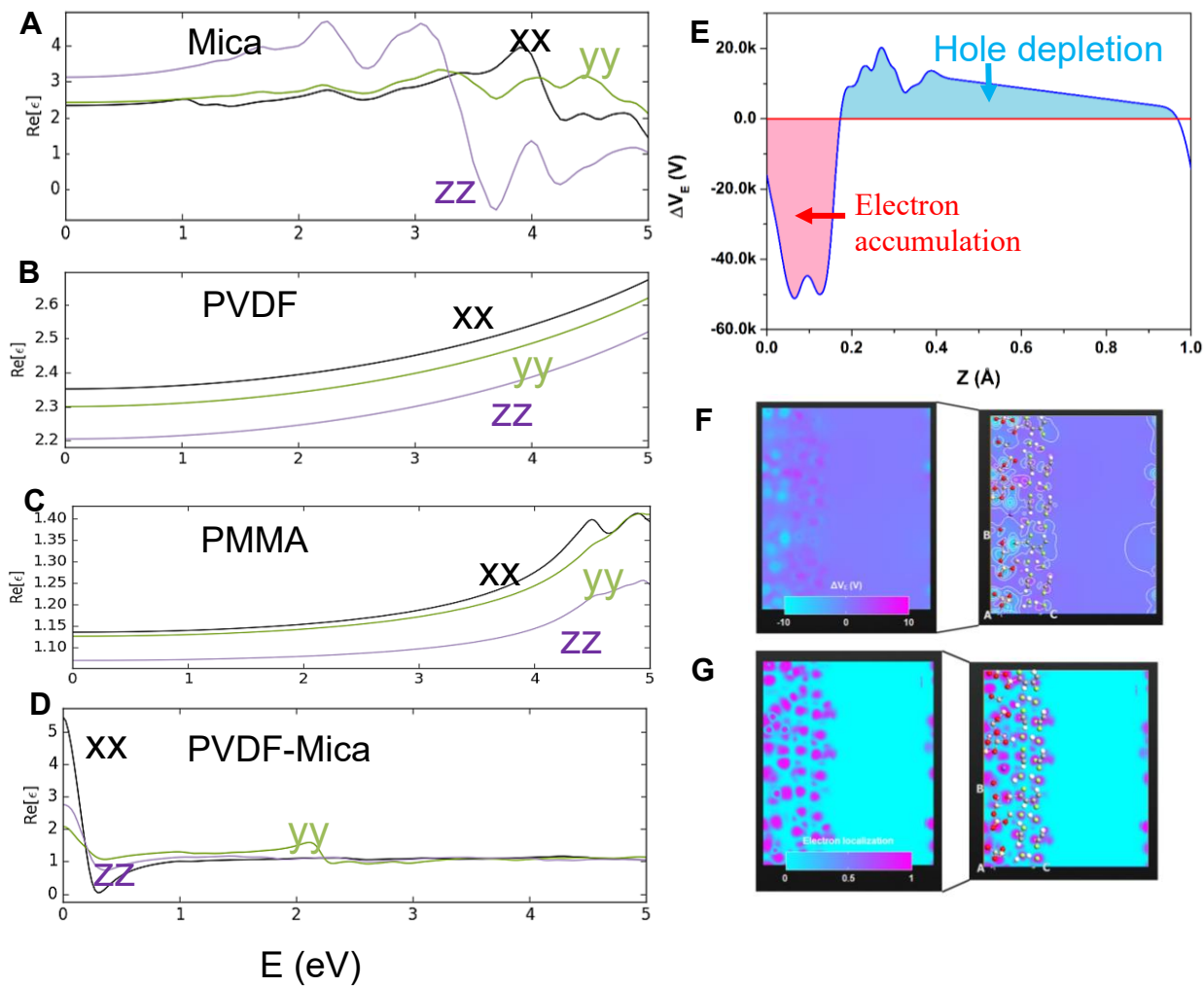

**Fig. 4. Computational evidence of permittivity enhancement in polymer-mica nanocomposites using density functional calculations.** (A) to (D) Real part of dielectric constant (ε) for Mica, PVDF, PMMA, and PVDF-Mica heterostructure respectively showing higher static ε for PVDF/Mica heterostructure, compared to their constituent polymers and Mica. (E) Plane-average of the electrostatic potential ($\Delta V_E$ (V)) along the normal direction of the Mica-PVDF interface. (F) The electrostatic difference potential (EDP) along the normal direction of the PVDF-Mica interface with contour plot (right) shows the spatial distribution of charge density across the interface (G) Electron localization function at the interface with the projection of atoms (right) showing intermolecular interactions. xx, yy, and zz are the dielectric tensor components along the x, y, and z-directions.

**Dielectric Strength and Capacitive Energy Density Measurements for Heterostructures**

The 2D inorganic nanofillers have been shown to enhance the dielectric strength of polymeric nanocomposites depending on the concentration of fillers. Typically, high fractions (≳ 10 wt. %) of randomly distributed 2D nanofillers are used to increase the dielectric strength (≈ 60-70 % enhancements) of polymeric nanocomposites [10]. Here, although the fraction of mica (or *h*-BN) nanofillers is very low (≈ 1 vol %), we expect the nanofillers to increase the dielectric strength significantly as the orientation of the 2D surface of the nanofillers are completely perpendicular to the applied electric field and can resist electrical treeing as shown in Figure 5 A. Indeed, Vogelsang and coworkers have demonstrated that the barriers oriented perpendicular to applied electric fields stop and perturb the propagation of electrical trees in bulk epoxy-Mica composites[23]. Furthermore, the above theoretical studies have confirmed that the in-plane alignment of the polymer chain axis leads to in-plane dipoles which helps enhance the dielectric strength or reduce the electrical treeing process [13]. Figure 5B shows two-parameter Weibull probability plots and fits of reference vs. stratified 2D Mica containing heterostructures capacitors measured at room temperature. The oriented Mica-based heterostructures capacitor shows an ultra-high dielectric strength of ≈ 1296 MV/m, which is ≈142% higher than the reference heterostructure capacitor using the thickness of the capacitor as 600 nm. This enhancement in dielectric strength is significant given that the nanofiller fraction is only ≈ 1 vol%. Simulation results predicted the 40% enhancement of the breakdown voltage in *h*-BN filler PVDF-CTFE and PVDF-TrFE-CFE composites at 8-10 vol% of filler concentration [*10*]. Along with increased dielectric strength, the β values increase from 7 for PVDF-PMMA-PVDF to 13 for PVDF-2D Mica/PMMA-PVDF heterostructure capacitor, showing that the reliability of dielectrics also increases. Figures 5C and 5D show the electrical displacement vs electric field and discharge energy density (J/cm$^3$) and efficiency for the stratified Mica-based heterostructures and the reference capacitors respectively. The stratified 2D Mica based polymer heterostructures show an ultra-high discharge energy density of ≈ 75 J/cm$^3$ with an efficiency of over 79% at a maximum electric field of 1247 MV/m, while the pure polymer heterostructures show a discharge energy density of 4.54 J/cm$^3$ with an efficiency of 78.9 % at an electric field of 483 MV/m. The energy density of the 2D Mica nanofiller-based capacitor is an order of magnitude higher than the pure polymer samples and stems from the increase in both permittivity and dielectric strength. The electrical treeing effect is significantly reduced in 2D-Mica interfaced capacitors due to the physical block of the electrical breakdown pathway by the 2D nanofillers in conjunction with the perpendicular alignment induced interfacial dipoles with respect to the field direction at the interface between polymer and fillers that oppose the applied E-field. This strategy results in the energy density of the PVDF/PMMA-2D Mica/PVDF samples is the highest to date for any thin film polymer composites as shown in SI Figures S16 (A) and S16 (B). Here we like to highlight that the literature capacitors shown in the comparison plots are based on free-standing films, while our 2D stratified nanofiller based measurements are substrate supported. We have used substrate supported films with relatively lower thickness (~500 nm) to maximize the effects of the nanosheets given that nanosheets have ~20 nm thickness. We expect that the same effects will hold true for thicker free-standing films with similar nanofiller density.

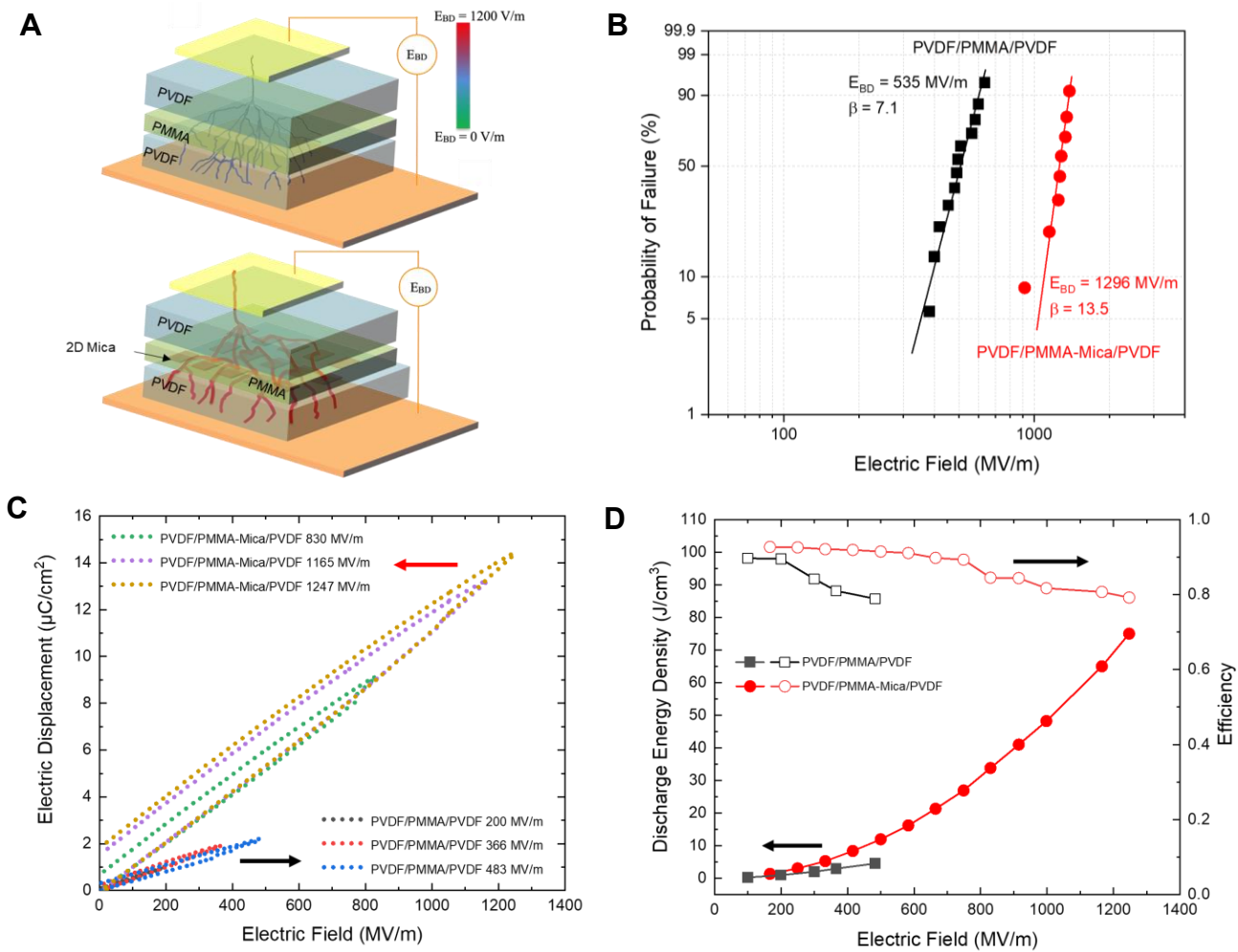

**Fig. 5. Ultra-high dielectric strength and capacitive energy storage properties of polymer heterostructures with oriented 2D Mica nanosheets.** (A) Schematic demonstrating enhanced resistance to electrical tree propagation by oriented 2D mica nanosheets in polymeric heterostructures. (B) Weibull dielectric strength of PVDF/PMMA/PVDF heterostructures and PVDF/PMMA-2D Mica/PVDF heterostructures. (C) Displacement vs electric field measurements and (D) discharge energy density and efficiency as a function of the applied electric field showing an ultra-high energy density of 75 J/cm$^3$ with efficiency over 79% at 1247 MV/m electric field for the Mica containing heterostructures.

**Versatility of Capacitive Energy Density Enhancement using Oriented 2D Nanofillers**

To test the versatility of ultra-high energy storage properties of stratified 2D nanofiller-based polymeric dielectrics, we further studied another two-dimensional dielectric material hexagonal Boron Nitride ($h$-BN) nanofillers in place of 2D Mica nanofillers. Figure 6A shows the enhancement of the dielectric constant of PVDF/PMMA-$h$-BN/PVDF heterostructures as compared to PVDF/PMMA/PVDF heterostructures (SI Fig. S8-S10). The $h$-BN based heterostructures show an increase of 40-45% in the permittivity at room temperature. The enhancement of the dielectric constant in PVDF/PMMA-$h$-BN/PVDF capacitor is lower than that observed for PVDF/PMMA-2D Mica/PVDF heterostructures, which might be attributed to the higher dielectric constant of Mica fillers compared to the $h$-BN. Dielectric permittivity $\varepsilon \propto N/\sqrt{E_g}$, $N$ is the density of the atom in the material, and $E_g$ is the energy gap [24]. Mica has a higher density (~3g/cm$^3$) compared to the $h$-BN (~2g/cm$^3$). Mica nanosheets have a lower band gap (~3 - 4 eV for 1-5 layers) [25] compared to $h$-BN (~6 eV) [26] making the dielectric constant of the Mica interfaced capacitor presumably higher than the $h$-BN interfaced capacitor. Figure 6B shows the Weibull probability plots of PVDF/PMMA/PVDF and PVDF/PMMA-$h$-BN/PVDF heterostructures. The dielectric strength of PVDF/PMMA-$h$-BN/PMMA heterostructures is ≈ 136% higher than that of the control heterostructures and is higher than that observed for their Mica counterparts. Analogous to Mica nanosheets, the high dielectric strength of $h$-BN based heterostructures is coupled to the high breakdown resistance of h-BN nanosheets, which is dependent on the 2D flake thickness[27]. Figures 6C and 6D show the electric displacement vs electric field and discharge energy density and efficiency plots of PVDF/PMMA-$h$-BN/PVDF and PVDF/PMMA/PVDF heterostructures. The $h$-BN based heterostructures can store an ultra-high energy density of ≈ 50 J/cm$^3$ with an efficiency of over 52% despite very low filler fraction (≈ 1 vol%), which is the highest among any literature reported polymeric dielectrics, except the 2D Mica nanosheets based heterostructures shown here. The lower value of energy density in *h*-BN based heterostructures is due to their slightly lower permittivities (polarizability and higher bandgap) as compared to the Mica heterostructures [28]. This demonstrates that the approach of using stratified 2D nanofillers within the polymer matrix is versatile for designing ultra-high energy density polymeric dielectric capacitors.

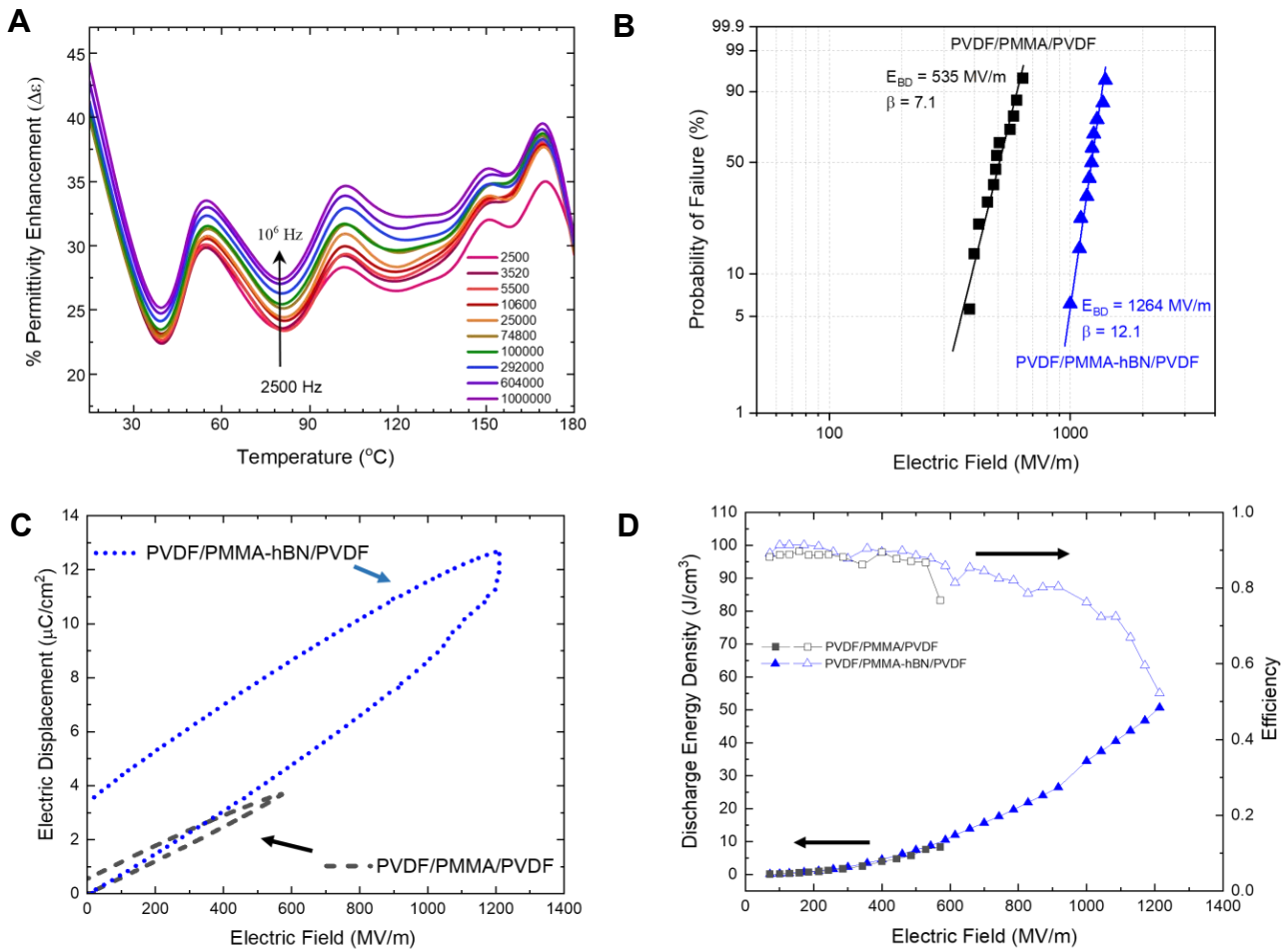

**Fig. 6. Ultra-high energy density in oriented *h*-BN based polymer heterostructures.** (A) Permittivity enhancement ($\Delta\varepsilon$) in PVDF/PMMA-*h*-BN/PVDF heterostructures as compared to PVDF/PMMA/PVDF control heterostructures. (B) Ultra-high dielectric strength of oriented PVDF/PMMA-*h*-BN/PVDF

heterostructures. (C) Electric displacement vs electric field and (D) Discharge energy density and efficiency of pristine and *h*-BN based heterostructures showing an ultra-high energy density of 50.7 J/cm$^3$ with an efficiency of 52% for *h*-BN based heterostructures using the thickness of the capacitor 750 nm measured using AFM (SI Fig. S8).

**Conclusions**

In conclusion, we envision that controlling the orientation of 2D nanofillers in polymeric nanocomposite-based dielectric films, such that the 2D nanofillers are perpendicular to the electrodes or applied electric field will be a rapidly growing field of de-novo research, given how significantly it increases the permittivity and electrical breakdown of the resulting nanocomposite films in our study. We use DFT calculations to verify our enhanced permittivity findings. Furthermore, the dielectric strength of polymer nanocomposite films increases by more than 130% as compared to the pristine polymeric films due to the "barrier effect" of the 2D nanomaterials. As a result, we develop polymer nanocomposites with energy density as high as ≈ 75 J/cm$^3$ with an efficiency of over 79% using PVDF/PMMA-2D Mica/PVDF heterostructures. We demonstrate that the use of orientation-controlled 2D nanofillers to boost the energy density of dielectric capacitors is versatile by testing PVDF/PMMA-*h*-BN/PVDF heterostructures as well. The intrinsic nanofiller permittivity and band gap are important considerations in this regard. We expect that the parallel orientation of 2D nanofillers with respect to the electrodes of the capacitor can be used for enhanced mechanical strength and efficient in-plane heat dissipation generated by pulsed power by using desired nanofillers like *h*-BN, especially important as these migrate into the power capacitor regime. Furthermore, this approach can be scaled up by using suitable large-scale fabrication of 2D materials using CVD and the printing technique of heterostructure capacitor fabrication. The capacitor could be fabricated on a flexible substrate like 2D materials and devices (graphene and 2D semiconductors) for potential application for flexible electronics. Finally, as an outlook, our study also inspires a paradigm shift in how to think about this promising field of aligned 2D nanofiller/polymer multilayer devices.

**Materials and Methods**

*Heterostructure device fabrication*

Poly(vinylidene fluoride) (PVDF) (MW: 534,000 by GPC), Dimethyl formamide (DMF), and Potassium hydroxide (KOH) were purchased from Sigma-Aldrich. PMMA was purchased from Microchem (currently named Kayaku Advanced Materials, Inc). Bulk 2D Mica sheets are purchased from SPI supplies. PVDF was dissolved in DMF, typically at 10 wt %. followed by stirring at 70 $^o$C temperature for 3 hours. This PVDF solution was used to make the thin film of PVDF using spin coating. The thin film of PVDF was prepared on ITO (Indium Tin Oxide) coated glass substrate. Each layer of PVDF thin film is prepared by spin coating of 10 wt % of PVDF solution in DMF solvent at 1700 rpm for 60 secs followed by annealing at 100 $^o$C for 5 minutes. To make PVDF film smoother and uniform, two layers (two times coating) of PVDF were coated on the ITO substrate.

*Dielectric permittivity, breakdown strength, and polarization measurements*

Temperature-dependent dielectric measurements were carried out using an LCR meter, Model: IM-3536 LCR Meter at the ambient condition with a temperature-controlled hot plate as a function of frequency range from 100 Hz to 8 MHz from room temperature (RT) to 180$^o$ C at ambient condition. All the samples were annealed initially at 180$^o$ C for 45 minutes before measuring frequency and temperature-dependent dielectric properties. Data were collected at each 10º C difference in the temperature range. The measured dielectric constant as a function of frequency for a sample (sample #S6) is given in Supplementary Information Figure S4. Dielectric constant (ε) was extracted using the relation $C = \frac{\varepsilon_0 \times \varepsilon \times A}{d}$, where $C$ is the capacitance, $\varepsilon_o$ is the permittivity in free space, $\varepsilon$ is the dielectric constant of the sample, $A$ is the area of the capacitor, and $d$ is the separation distance between two plates or thickness of dielectric materials (heterostructure capacitor). Our capacitor area for dielectric measurements is typically (~350 × 350 µm$^2$) area, fabricated using conducting Ag paint or thermally deposited Au electrodes on the top PVDF layer (top electrode) (Supporting Information Fig S5). The sample thickness measured from contact mode AFM for the samples presented in Fig. 2 (for Mica) was 500+/-50 nm. The average thickness was used to extract the dielectric constants. The thickness of $h$-BN interfaced heterostructure capacitors was measured between 600+/-50 nm.

Breakdown voltage measurements were conducted using the PolyK testing system PK CPE-1801 equipped with a High-voltage TREK amplifier Model # 609D-6 using a wire loop-based contact for the top electrode to minimize mechanical damage to the films. A ramp rate of 20 V/s was selected for the testing and the breakdown values were recorded using the PolyK software. The data were analyzed using a two-parameter Weibull fit as

$$P(E) = 1 - \exp(-(E/E_{BD})^\beta),$$

where $E$ is the breakdown field, $E_{BD}$ is the dielectric breakdown strength, i.e., breakdown strength at 63.2% probability of failure, $\beta$ is the shape parameter and $P(E)$ is the probability of failure.

The electric displacement vs. electric field ($D$-$E$ loop) measurements were performed using the PolyK PK CPE-1801 system. The top electrodes for the $D$-$E$ loop measurement were fabricated using Gallium-Indium liquid metal and had diameters of around ~200-300 microns. The liquid metal electrodes were connected to the instruments using needlepoint contacts. Raman measurements on bulk Mica was done using a 785 nm laser source (MacroRAM™ benchtop Raman spectrometer) in the range of 100 to 1000 cm$^{-1}$ at ambient condition. Fourier Transform Infrared Spectroscopy (FTIR) measurement was conducted using Perkin–Elmer-Spectrum two FT-IR spectrometers in the range 4000–450 cm$^{-1}$. Atomic force microscope (AFM) images of the surface morphology and height measurements were conducted using Veeco Dimension 3100 AFM instrument in contact mode.

*Density Functional Theory Calculations*

The dielectric constant of polymers, Mica, and heterostructures was calculated using the DFT method. The optimized structure of the different geometry of polymer and mica and their heterostructure are presented in Fig. S12. We also extracted the refractive index of the Mica and polymer shown in the Supporting Information section (Fig. S13). The refractive indexes for Mica, PVDF, and PMMA were evaluated and found to be in good agreement with the previously experimentally reported values (29,30).

*Computational Details*

To estimate the optical properties of mica-polymer heterostructures, the energetics of bulk nanomaterials and their interface geometries are determined within the framework of density functional theory (DFT). The lowest energy structures of the pristine materials on the potential energy surface (PES) are identified by fully relaxing the unit cell structures employing the generalized gradient approximation (GGA)-based Perdew–Burke–Ernzerhof (PBE) [31] exchange-correlation functional and the double zeta polarized (DZP) basis sets in conjunction with the norm-conserving pseudopotentials for all the constituent atoms, and a mesh cut-off energy of 185 Hartree. The atomic positions of the unit cells are relaxed until the residual forces on each atom are less than 0.05 eV/Å. The fully relaxed structures of mica and PVDF/PMMA are then exploited to produce the mica-polymer interface geometry, as shown in Fig. S12. To identify the epitaxial interface where both the strain and the area of the coincidence interface cell are minimized, we have implemented the lattice matching method [32] that probes the interface compatibility by incorporating convenient effects on the lattice constant in the direction perpendicular to the interface plane i.e., the Poisson effect. The predicted low-strain epitaxial interface, with a mean strain of 0.2%, is further subjected to relaxation by DFT based method. The k-points in the Monkhorst-Pack grid are set to (4 x 4 x 4) and (4 x 4 x 1) for the relaxation of single-phase supercells and interface cells, respectively. Furthermore, a vacuum spacing of 20 Å perpendicular to the lattice plane was used during the structural relaxation to overcome the spurious interactions between repeated images. All DFT computations were performed using the QuantumWise ATK software package [33].

Next, to explore the impact of polymer matrix deposition on the optical properties of the mica-polymer heterostructures, the optical functions of the pristine and heterostructure materials are derived from the DFT computations within the framework of Kubo-Greenwood formalism. The Kubo-Greenwood formula can be used to compute the susceptibility tensor as follows [34, 35]

$$\chi_{ij}(\omega) = -\frac{e^2 \hbar^4}{m^2 \epsilon_0 V \omega^2} \sum_{nm} \frac{f(E_m) - f(E_n)}{E_{nm} - \hbar\omega - i\Gamma} \pi_{nm}^i \pi_{mn}^j$$

Where $\pi_{nm}^i$ is the $i$-th component of the dipole matrix element between state $n$ and $m$, $f(E_{m/n})$ is the Fermi function calculated at the band energy, $\Gamma$ is the energy broadening, and $V$ is the volume. The frequency-dependent complex dielectric function $\varepsilon(\omega) = \varepsilon_1(\omega) + i\varepsilon_2(\omega)$ is associated with the susceptibility as $\varepsilon(\omega) = 1 + \chi(\omega)$, which represents the linear response of the dielectric properties of the material.

The refractive index $n$ is interrelated to the complex dielectric constant through

$$n + ik = \sqrt{\varepsilon}$$

where $k$ is the extinction coefficient. The values of $n$ and $k$ could be obtained from the real ($\varepsilon_1$) and imaginary ($\varepsilon_2$) parts of the dielectric constant by using the following relations

$$n = \sqrt{\frac{\sqrt{\varepsilon_1^2 + \varepsilon_2^2} + \varepsilon_1}{2}}$$

$$k = \sqrt{\frac{\sqrt{\varepsilon_1^2 + \varepsilon_2^2} - \varepsilon_1}{2}}$$

The optical absorption coefficient is related to the extinction coefficient through the following relation [36]:

$$\alpha_a = 2\frac{\omega}{c}k$$

**Acknowledgments**:

This work was supported by NSF HBCU-UP Excellence in Research NSF-DMR-1900692 and DoD Grant # G634E27. NRP acknowledged NSF-PREM through NSF-DMR-1826886 for partial support. NRP and SB acknowledged the support from Princeton Alliance for Collaborative Research and Innovation (PACRI) Grant # PACRI-JSU-02.

**Data availability:** The data that support the findings of this study are available from the authors on reasonable request.

**Competing Interests:**

The author declares no competing interests.

**Author Contribution:** N.R.P. conceived the project and work performed based on ideas of N.R.P., A.K. M.S, and D.R. P.D. S.B. and N.R.P. fabricated the heterostructure capacitor devices, Q.Z. and Q.D. deposited the Au contacts, P.D. B.S. R.T. and N.R.P. performed the temperature-dependent dielectric measurements and analyzed the corresponding data with B.B., P.D., A.P., and P.R. have performed Raman and FTIR measurements, M.S., R.N., D.R., and A.K. performed and analyzed the breakdown voltage and P-E loop measurements. P.D. and N.R.P. performed the AFM characterizations on the flakes. N.R.P. A.K. M.S. wrote the manuscript with the input of all the co-authors. All authors contributed to the scientific discussion and editing of the manuscript.

**Additional Information:**

**Supplementary Information** containing Raman spectra, FTIR spectra, AFM measurements, microscopy images, additional DFT simulations and dielectric measurements, and a comparison of our devices with literature capacitors is attached.